\documentstyle[epsfig,aps,preprint]{revtex}
\tighten
\begin{document}
\draft

\title{Composite antisymmetric-tensor Nambu-Goldstone bosons 
in a four-fermion interaction model and the Higgs mechanism}
\author{V. Dmitra\v sinovi\' c}
\address{Department of Physics and Astronomy,\\
University of South Carolina, Columbia, SC 29208, USA\\
e-mail: dmitra@nuc003.psc.sc.edu}
\date{\today}
\maketitle
\begin{abstract}
Starting from the Fierz transform of the two-flavour 't Hooft interaction
(a four-fermion Lagrangian with antisymmetric Lorentz 
tensor interaction terms augmented by an NJL type Lorentz scalar
interaction responsible for dynamical symmetry breaking and quark mass 
generation), we show that: (1)  
antisymmetric tensor Nambu-Goldstone bosons appear provided 
that the scalar and tensor couplings stand in the proportion of two 
to one, which ratio appears naturally in the Fierz transform of the 
two-flavour 't Hooft interaction; 
(2) 
non-Abelian vector gauge bosons coupled to this system 
acquire a non-zero mass.
Axial-vector fields do not mix with antisymmetric tensor fields, so there 
is no mass shift there.
\end{abstract}
\pacs{PACS numbers: 11.15.Ex, 11.30.Qc, 11.10.St}

\section{Introduction}

It has long been known that four-fermion contact interactions of the 
Nambu and Jona-Lasinio [NJL] type
can lead to dynamical symmetry breaking along with associated composite 
spinless
Nambu-Goldstone [NG] bosons \cite{njl61}. Such interactions have been extended
to include all but six of the 16 independent Dirac matrix bilinears. The six 
still
unexplored terms correspond to the antisymmetric [a.s.] tensor 
$\sigma_{\mu \nu} = (i/2) \left[\gamma_{\mu}, \gamma_{\nu}\right]$ 
self-interaction, which leads 
after bosonisation to antisymmetric tensor bosonic excitations \cite{es74}.
Elementary antisymmetric tensor fields were introduced into field theory
by Ogievetskii and Polubarinov \cite{op66}, into string theory
by Kalb and Ramond \cite{kr74} and into models of confinement by Nambu 
\cite{nambu74}, but their origin was not specified.

The purpose of this paper is to show that: 1) a model that combines 
the NJL dynamical symmetry breaking with an antisymmetric tensor quark 
self-interaction, obtained from the two-flavour 't Hooft interaction by a 
Fierz rearrangement, leads to massless
composite antisymmetric tensor and pseudotensor NG bosons at a.s. 
tensor coupling equal to one half of 
the scalar coupling constant $G_T = {1\over 2} G_S$; 2) 
vector boson coupled to this system acquires a non-zero mass. 

\section{Preliminaries}

We shall work with a chirally symmetric field theory described by 
\begin{eqnarray} 
{\cal L}_{\rm ENJL} = \bar{\psi} \big[{\rm i} {\partial{\mkern -10.mu}{/}} ]\psi
&+&
G_{S} \Big[ (\bar{\psi} \psi)^2 + 
(\bar{\psi} {\rm i} \gamma_5 \mbox{\boldmath$\tau$} \psi)^2 \Big] 
\nonumber \\ 
&+& 
G_{T} \Big[ (\bar{\psi} \sigma_{\mu \nu} 
\psi)^2 + 
(\bar{\psi} i\gamma_5 \sigma_{\mu \nu} \mbox{\boldmath$\tau$} \psi)^2 \Big]
\: , 
\label{e:lag1}  
\end{eqnarray} 
where $\psi$ is a two-component (isospinor) Dirac field.
We shall have no need for the pseudoscalar term in the first line of 
Eq. (\ref{e:lag1}) in the forthcoming analysis, but keep it to preserve 
the underlying $SU_{L}(2) \times SU_{R}(2)$ chiral symmetry. 
The antisymmetric tensor 
self-interaction in the second line of Eq. (\ref{e:lag1})
preserves this chiral symmetry,
but violates the axial baryon number $U_{A}(1)$ conservation \cite{vd97}. 
[This self-interaction is related to the two-flavour 't Hooft interaction 
\cite{th76} by a Fierz identity.] 

There is another, hidden symmetry of the a.s. tensor term
(the second line in the Lagrangian (\ref{e:lag1}) that has not been
explored in this context heretofore and which we shall call
the ``duality symmetry''. It is induced by 
the identity 
\begin{eqnarray} 
\gamma_{5} \sigma_{\mu \nu} 
&=& 
{i \over 2} \varepsilon_{\mu \nu \alpha \beta}\sigma^{\alpha \beta} 
= i \sigma_{\mu \nu}^{\star} = i {\tilde \sigma}_{\mu \nu}
\label{e:ident}
\end{eqnarray}
which allows
the second line in the Lagrangian (\ref{e:lag1}) to be written as
\begin{eqnarray} 
G_{\rm T} \Big[\left(1 + \lambda \right) 
(\bar{\psi} \sigma_{\mu \nu} \psi)^2 + 
\lambda  
(\bar{\psi} i\gamma_5 \sigma_{\mu \nu} \mbox{\boldmath$\tau$} \psi)^2 
+ \lambda (\bar{\psi} i \gamma_5 \sigma_{\mu \nu} \psi)^2 + 
\left(1 + \lambda \right)
(\bar{\psi} \sigma_{\mu \nu} \mbox{\boldmath$\tau$} \psi)^2 \Big]
\: . 
\label{e:lag3}  
\end{eqnarray} 
where $\lambda$ is an arbitrary (real) ``duality-symmetry gauge 
fixing parameter''. 
Manifestly, physical predictions of this model must be independent of 
$\lambda$. 
As a particular consequence of duality the a.s. tensor term in the Lagrangian 
(\ref{e:lag1}) 
vanishes identically in the Abelian, i.e. one-flavour ($N_f = 1$) case:
\begin{eqnarray} 
G \Big[ (\bar{\psi} \sigma_{\mu \nu} \psi)^2 + 
(\bar{\psi} i \gamma_5 \sigma_{\mu \nu} \psi)^2 \Big] &=& 0\: . 
\label{e:lag2}  
\end{eqnarray} 
We shall have to pay particular attention to maintaining the duality-symmetry 
of this theory in the approximations used here.

\begin{figure}
\begin{center}
\epsfig{file=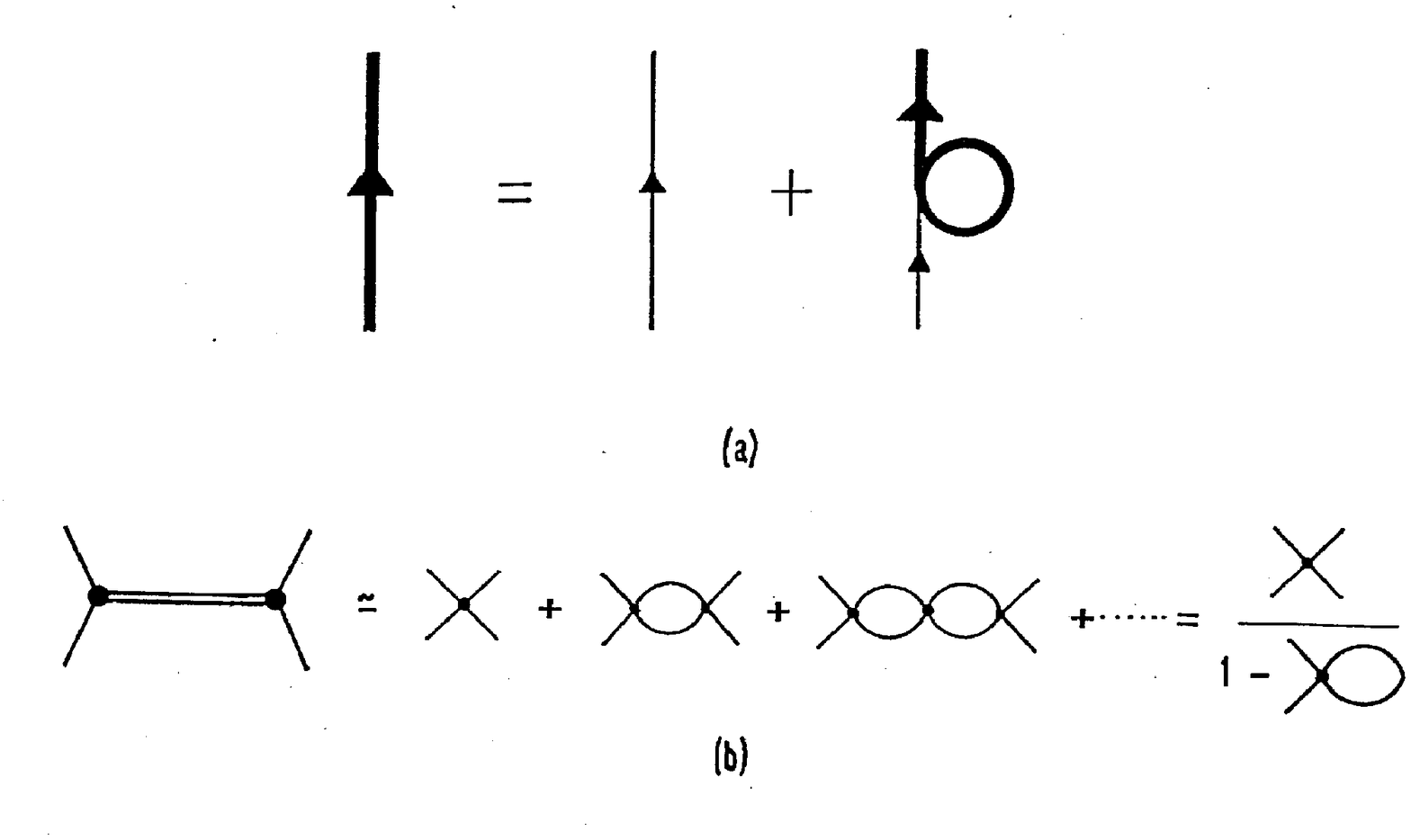,width=16cm} 
\end{center} 
\caption{
The two Schwinger-Dyson equations determining the dynamics of this model
in the Hartree + RPA approximation:
(a) the one-body,  or ``gap'' equation; and (b) the two-body, or Bethe-Salpeter 
equation. The thin solid line is the bare, or ``current'' quark, the heavy solid
line is the constituent quark and the double solid line is either the composite
antisymmetric tensor-, or pseudo-tensor boson.}
\label{f:1}
\end{figure}

The non-perturbative dynamics of the model, to
leading order in $1/N_C$, are described by two Schwinger-Dyson [SD]
equations: (i) the gap equation,  Fig.~\ref{f:1}.a; and (ii) the 
Bethe-Salpeter [BS] equation,  Fig.~\ref{f:1}.b. Our model
has three parameters: two positive coupling constants $G_S , G_T$ of dimension 
(mass)$^{-2}$ and a regulating cut-off $\Lambda$ that determines the mass 
scale.
In the Hartree approximation, i.e., to leading order in $1/N_C$, the gap 
equation 
\begin{eqnarray} 
m  &=& 16 i N_{C} G_{S} \left\{ \int {d^{4} p \over {(2 \pi)^{4}}} 
{m \over {p^{2} - m^{2}}}\right\}_{\rm reg}  \ ~,
\label{e:gap}
\end{eqnarray} 
for 
$N_{c}$ colours, regulated
following either Pauli and Villars (PV), or dimensionally \cite{iz80},
establishes a relation between the constituent quark mass $m$ and the free 
parameters $G_S$ and $\Lambda$. 
[Regularization of the quantity in braces is indicated by its subscript.]
This relation is not one-to-one, however: there is a double continuum of 
allowed
$G_S$ and $\Lambda$ values that yield the same nontrivial solution $m$ to the 
gap equation. One of these 
degeneracies can be eliminated by fixing the spinless ($0^-$) Nambu-Goldstone 
[NG] boson decay constant $f_{p}$ at its observed value.

\section{Bethe-Salpeter equation: antisymmetric polarization tensors}

The second Schwinger-Dyson equation is an 
inhomogeneous Bethe-Salpeter (BS) equation 
\begin{eqnarray}
{\bf D} = 2 {\bf G} + 2 {\bf G\, \Pi\, D},
\label{e:bse}
\end{eqnarray}
describing the scattering of quarks and antiquarks,  Fig.~\ref{f:1}.b. 
Because there is potential for mixing between the channels
all the objects in the BS equation are 4 $\times$ 4 matrices.
Here ${\bf G}$ is the (effective) T coupling constant matrix
\begin{eqnarray}
{\bf G} &=& 
G_T 
\left(\begin{array}{cc}
\alpha {\bf 1} & 0 \\
0 &  \beta {\bf 1}
\end{array}\right)~
\label{e:G}
\end{eqnarray} 
corresponding to the interaction Lagrangian
\begin{eqnarray} 
G_T \Big[\alpha  (\bar{\psi} \sigma_{\mu \nu} \psi)^2 + \beta 
(\bar{\psi} i \gamma_5 \sigma_{\mu \nu} \psi)^2 \Big] &=& 0\: , 
\label{e:lag4}  
\end{eqnarray} 
parametrized by a couple of parameters $\alpha, \beta$ 
in each isospin channel.
Here ${\bf 1}$ is a 2 $\times$ 2 unit matrix (corresponding to two
channels), the ``upper'' submatrix   
describes the T and the ``lower'' one the PT channel,
and ${\bf \Pi}$ is the a.s. polarization tensor matrix whose matrix elements 
we must evaluate. 

To leading order in $N_C$, the polarization $\Pi (k)$ is just a single-loop 
diagram. The form of the interaction in Eq. (\ref{e:lag1}) gives rise 
to scattering in four channels: the familiar isovector-pseudoscalar (pion) 
channel 
and the isoscalar-scalar (sigma) channel and the corresponding two 
antisymmetric (pseudo-)tensor channels. 

\paragraph*{Antisymmetric polarization tensors}
To leading order in $N_C$, the polarization $\Pi (k)$ is just a single-loop 
diagram. The form of the interaction in Eq. (\ref{e:lag1}) gives rise 
to four polarization functions: 
two (isovector and isoscalar) 
in the antisymmetric tensor and pseudo-tensor channels each.
We start with the a.s. tensor polarization:
\begin{eqnarray}
-i\Pi^{{\rm T},ab}_{\mu\nu;\alpha\beta} &=& 
-i\Pi^{\rm T}_{\mu\nu;\alpha\beta}(q^2) \delta^{ab}=
2 N_C \delta^{ab}
\int\frac{d^4p}{(2\pi)^4}{\rm tr}[i \sigma_{\mu \nu}
S(p+q) i \sigma_{\alpha\beta}
S(p)] \
\label{e:tpol}
\end{eqnarray}
and similarly for the pseudotensor polarization
\begin{eqnarray}
-i\Pi^{{\rm PT},ab}_{\mu\nu;\alpha\beta} &=& 
-i\Pi^{\rm PT}_{\mu\nu;\alpha\beta}(q^2) \delta^{ab}=
2 N_C \delta^{ab}
\int\frac{d^4p}{(2\pi)^4}{\rm tr}[\gamma_{5} \sigma_{\mu \nu} 
S(p+q) \gamma_{5} \sigma_{\alpha\beta} S(p)]~. \
\label{e:ptpol}
\end{eqnarray}
There are two independent a.s. tensors, 
\begin{eqnarray}
T_{\mu \nu, \alpha \beta}^{\rm I} &=& {1 \over 2}
\left[g_{\mu \alpha} g_{\nu \beta} - g_{\mu \beta} g_{\nu \alpha}\right]  
\nonumber \\
T_{\mu \nu, \alpha \beta}^{\rm II} &=& {1 \over 2 q^{2}}
\left[g_{\mu \alpha} q_{\nu} q_{\beta} + g_{\nu \beta} q_{\mu} q_{\alpha} 
- g_{\mu \beta} q_{\nu} q_{\alpha} - g_{\nu \alpha} q_{\mu} q_{\beta} \right]~,
\label{e:tens1}
\end{eqnarray}
so we may write the polarization tensors as follows
\begin{eqnarray}
\Pi_{\mu \nu, \alpha \beta}^{\rm PT}(q^2 ) &=& 
\Pi_{\rm PT}^{\rm I}(q^2 ) T_{\mu \nu, \alpha \beta}^{\rm I} +
\Pi_{\rm PT}^{\rm II}(q^2 ) T_{\mu \nu, \alpha \beta}^{\rm II}
\nonumber \\
\Pi_{\mu \nu, \alpha \beta}^{\rm T}(q^2 ) &=& 
\Pi_{\rm T}^{\rm I}(q^2 ) T_{\mu \nu, \alpha \beta}^{\rm I} +
\Pi_{\rm T}^{\rm II}(q^2 ) T_{\mu \nu, \alpha \beta}^{\rm II},.
\label{e:pi}
\end{eqnarray}
without loss of generality. In four dimensions the identity 
(\ref{e:ident}) demands the following ``duality'' relations 
between the tensor and pseudotensor polarizations
\begin{eqnarray}
{\tilde \Pi}_{\mu \nu, \alpha \beta}^{\rm T}(q^2) &=& 
{1 \over 4} \varepsilon_{\mu \nu \gamma \delta}
\Pi^{\gamma \delta, \sigma \rho}_{\rm T}(q^2 )
\varepsilon_{\sigma \rho \alpha \beta}
\nonumber \\
&=& 
^{\star} \Pi_{\mu \nu, \alpha \beta}^{{\rm T}~ \star}(q^2 )
= \Pi_{\mu \nu, \alpha \beta}^{\rm PT}(q^2 )  
\nonumber \\
{\tilde \Pi}_{\mu \nu, \alpha \beta}^{\rm PT}(q^2 ) &=& 
{1 \over 4} \varepsilon_{\mu \nu \gamma \delta}
\Pi^{\gamma \delta, \sigma \rho}_{\rm PT}(q^2 )
\varepsilon_{\sigma \rho \alpha \beta}
\nonumber \\
&=& 
^{\star} \Pi_{\mu \nu, \alpha \beta}^{{\rm PT}~ \star}(q^2 )
= \Pi_{\mu \nu, \alpha \beta}^{\rm T}(q^2 ) .
\label{e:dual}
\end{eqnarray}

Equations (\ref{e:dual}) imply the following constraints
\begin{eqnarray}
\Pi_{\rm T}^{\rm I} &=& - 
\left(\Pi_{\rm PT}^{\rm I} + \Pi_{\rm PT}^{\rm II}\right)
\nonumber \\
\Pi_{\rm T}^{\rm II} &=& \Pi_{\rm PT}^{\rm II} = \Pi^{\rm II} ~.
\label{e:dual3}
\end{eqnarray}
These results are both theoretically important and valuable in the evaluation 
of the polarization functions.

We evaluate the traces in four dimensions, so as to avoid the ambiguities
of the definition of $\gamma_5$ matrix in non-integer dimensions and 
to conform with the duality requirements (\ref{e:dual3}). One finds
\begin{eqnarray}
\Pi^{\rm I}_{\rm PT}(s) &=& - 
{1 \over 3} \left[- 16 f_{p}^{2} - 48 i N_C J
+ 2 g_{p}^{-2} \left(s + 8 m^2\right) F(s) \right]
= \Pi^{\rm I}_{\rm T}(s) - 8 f_{p}^{2} F(s)
\nonumber \\
\Pi^{\rm I}_{\rm T}(s) &=& - 
{1 \over 3} \left[
- 16 f_{p}^{2} - 48 i N_C J
+ 2 g_{p}^{-2} \left(s - 4 m^2\right) F(s) \right]
\nonumber \\
\Pi^{\rm II}(s) &=& 
{4 \over 3} g_{p}^{-2} \left[ 
s  F(s) + 2 m^2 \left(F(s) - 1\right) \right] = - 2 \Pi_{\rm V}(s) ~,
\label{e:pi1p}
\end{eqnarray}
where 
\begin{mathletters}
\begin{eqnarray}
F(s) &=& 
{\left\{I(s)\right\}_{\rm reg} \over{\left\{I(0)\right\}_{\rm reg}}} 
= 1 - \frac{3 g_{p}^2}{2 \pi^2} \{\sqrt{- f}
{\rm Arccot}{\sqrt{-f}} - 1 \}_{\rm reg}
\label{e:ff} \\
f &=& 1 - 4m^2/s 
\label{e:f} \\
g_{p}^{-2} &=& 
\left({f_{p} \over{m}}\right)^{2}
= - 4 i N_C \left\{I(0)\right\}_{\rm reg} ~,
\label{e:gpi}\
\end{eqnarray}
and $I(k), J$ are the, respectively, logarithmically and quadratically 
divergent one-loop integrals 
\begin{eqnarray} 
I(k^2) &=&  \int {d^{4} p \over {(2 \pi)^{4}}}
{1 \over {[p^{2} - m^{2}][(p + k)^{2} - m^{2}]}} 
\label{I(k)} \\
J &=&  \int {d^{4} p \over {(2 \pi)^{4}}}
{1 \over {[p^{2} - m^{2}]}} ~.
\label{J}
\end{eqnarray}  
\end{mathletters}
Eq. (\ref{e:gpi}) describes the Goldberger-Treiman [GT] relation, which is
a chiral Ward identity and happens to hold in most regularization schemes,
even when they violate other Ward identities. 
The
set of a.s. tensor polarization functions (\ref{e:pi1p}) requires
\begin{equation}
m^2 \left\{I(0)\right\}_{\rm reg} = 
\left\{J\right\}_{\rm reg}
\label{e:igap}
\end{equation}
in order to satisfy the duality constraint Eq. (\ref{e:dual3}), however.
This is the same condition that converts the gauge-variant sharp 
Euclidean space cutoff vector polarization 
function $\Pi_{\rm V}$ into the gauge invariant [g.i.] one in Eq. 
(\ref{e:pi1p}).
This procedure eliminates the quadratically divergent integral from 
the vector polarization function $\Pi_{\rm V}(s)$, Ref. \cite{vd99}, 
and thus makes sure that the photon remains massless. Eq. (\ref{e:igap})
holds in the Pauli and Villars (PV) scheme for only one value of the cutoff 
$\Lambda$, and 
even the signs of the two sides of this ``equation'' in the PV scheme 
coincide only in a 
narrow region of cutoff $\Lambda$ and mass $m$ values.
With dimensional regularization, however,  
\begin{equation}
m^2 \left\{ 
\int {d^{n} p \over {(2 \pi)^{n}}} {1 \over {\left(p^{2} - m^{2}\right)^{2}}}
\right\}_{\rm dim} =  
\left({n \over 2} - 1 \right)
\left\{\int {d^{n} p \over {(2 \pi)^{n}}} {1 \over {p^{2} - m^{2}}}
\right\}_{\rm dim}
\label{e:igap1}
\end{equation}
holds as an {\it identity}. With the help of the GT relation,
however, Eq. (\ref{e:igap}) can be written as 
\begin{equation}
4 \left\{f_{p}^{2}\right\}_{PV} = - {1 \over{G_{S}}}.
\label{e:cond}
\end{equation} 
As a consequence of dual symmetry we are facing here 
two alternatives: (1) 
imaginary decay constant $f_{p}$ and composite boson-fermion coupling 
constant $g_{p}$, or (2) negative four-fermion coupling constant $G_{\rm S}$.
We choose the latter. In other words, with g.i. and duality-invariant 
regularizations, such as the 
dimensional one, the sign of the
scalar coupling constant $G_{\rm S}$ in Eq. 
(\ref{e:lag1}) is opposite to the usual one. 
As pointed out earlier, this coupling constant is not an observable, so we
may flip its sign with impunity so long as observables, such as the fermion
mass and the p.s. decay constant $f_p$ remain unaffected, which is precisely 
the case here. 
This seems a small price to pay for a regularization scheme that is 
consistent with the gauge and duality symmetries. 
Moreover, this prescription also allows us to use the dimensional 
regularization in the NJL model.

Upon enforcing the relation (\ref{e:cond}) in Eqs. (\ref{e:pi1p}), we find
\begin{eqnarray}
\Pi^{\rm I}_{\rm PT}(s) &=& - 
{2 \over 3} g_{p}^{-2}  
\left[
\left(s + 6 m^2\right) F(s)
+ 2 m^2 \left(F(s) - 1\right) \right]
= \Pi^{\rm I}_{\rm T}(s) - 8 f_{p}^{2} F(s)
\nonumber \\
\Pi^{\rm I}_{\rm T}(s) &=& - 
{2 \over 3} g_{p}^{-2} \left[
\left(s - 6 m^2\right) F(s)
+ 2 m^2 \left(F(s) - 1\right) \right]
= \Pi_{\rm A}(s) 
\nonumber \\
\Pi^{\rm II}(s) &=& 
{4 \over 3} g_{p}^{-2} \left[ 
s  F(s) + 2 m^2 \left(F(s) - 1\right) \right] = - 2 \Pi_{\rm V}(s)~.
\label{e:pi2p}
\end{eqnarray}
We shall call these results ``gauge invariant a.s. tensor polarization 
functions''
and use them forthwith as the only option. 

\paragraph*{Antisymmetric tensor-pseudotensor mixing matrix elements}
The duality identity (\ref{e:ident}) also connects the a.s. polarization 
tensor 
to the a.s. polarization pseudo-tensor and hence to the mixing of the 
two modes via the nonvanishing PT-T transition matrix element [ME]
\begin{eqnarray}
\Pi^{{\rm PT - T}}_{\mu\nu;\alpha\beta} &=& 
2 i N_C \int\frac{d^4p}{(2\pi)^4}{\rm tr}[\gamma_5  \sigma_{\mu \nu}
S(p+q) i \sigma_{\alpha\beta} S(p)] 
\nonumber \\
&=& 
- {1 \over 2} \varepsilon_{\mu \nu}^{~~~\gamma \delta}
\Pi_{\gamma \delta, \alpha \beta}^{\rm T}(q^2 )
= - ^{\star} \Pi_{\mu \nu, \alpha \beta}^{\rm T}(q^2 )
\nonumber \\ &=& 
{1 \over 2} \varepsilon_{\alpha \beta}^{~~~\gamma \delta}
\Pi_{\mu \nu, \gamma \delta}^{\rm PT}(q^2 ) 
= \Pi_{\mu \nu, \alpha \beta}^{{\rm PT}~\star}(q^2 )\ .
\label{e:tpt}
\end{eqnarray}
This leads to the following product of matrices
\begin{eqnarray}
{\bf G \,\Pi} &=& {\bf \Pi \,G} =
G_T
\left(\begin{array}{cc}
\left(\alpha - \beta \right) \Pi_{\rm T} & 0 \\
0 & - \left(\alpha - \beta \right) \Pi_{\rm PT} 
\end{array}\right)~
\label{e:PiG}
\end{eqnarray} 
As we can see, the T and PT channels are separate now,
the only effect of mixing being the perhaps unexpected, yet 
duality-gauge-invariant 
linear combination $\left(\alpha - \beta \right) = 
\left(1 + \lambda \right) - \lambda = 1$.

\section{Bethe-Salpeter equation: antisymmetric tensor meson propagators}

After separating the two opposite parity channels in Eq. (\ref{e:bse})
we find as the solutions
\begin{eqnarray}
D_{\rm T}^{\rm I} &=& 
{2 \alpha G_T \over{
1 - 2 \left(\alpha - \beta \right) G_T \Pi_{\rm T}^{\rm I}(q)}} 
\nonumber \\
D_{\rm T}^{\rm II} &=&  
2 \alpha G_T \left[
{1 \over{1 - 2 \left(\alpha - \beta \right) G_T \left(\Pi_{\rm T}^{\rm I}(q)
+ \Pi^{\rm II}(q)\right)}} - 
{1 \over{ 1 - 2 \left(\alpha - \beta \right) G_T \Pi_{\rm T}^{\rm I}(q)}} 
\right]
\nonumber \\
&=& 
2 \alpha G_T \left[
{1 \over{1 + 2 \left(\alpha - \beta \right) G_T \Pi_{\rm PT}^{\rm I}(q)}} - 
{1 \over{ 1 - 2 \left(\alpha - \beta \right) G_T \Pi_{\rm T}^{\rm I}(q)}} 
\right]
\nonumber \\
D_{\rm PT}^{\rm I} &=& 
{2 \beta G_T \over{
1 + 2 \left(\alpha - \beta \right) G_T \Pi_{\rm PT}^{\rm I}(q)}} 
\nonumber \\
D_{\rm PT}^{\rm II} &=&  
2 \beta G_T \left[
{1 \over{1 - 2 \left(\alpha - \beta \right) G_T \Pi_{\rm T}^{\rm I}(q)}} - 
{1 \over{ 1 + 2 \left(\alpha - \beta \right) G_T \Pi_{\rm PT}^{\rm I}(q)}} 
\right]~,
\label{e:prop2}
\end{eqnarray}
Hence we see that the denominators are duality-gauge invariant, but not so the
numerators.

\paragraph*{Duality constraints on the solutions to the BS equation}

Now remember that a T channel propagator can be turned into 
a PT one by the duality transformation.
Hence the ``complete'' propagators 
are given by
\begin{eqnarray}
D_{\rm T}^{'}(G) &=& 
D_{\rm T}(G) + {\tilde D}_{\rm PT}(G) 
\nonumber \\
D_{\rm PT}^{'}(G) &=& 
D_{\rm PT}(G) + {\tilde D}_{\rm T}(G) 
~,
\label{e:dual7}
\end{eqnarray} 
respectively.
As we shall show below, these two complete
propagators are duality-gauge 
invariant and they describe two distinct particles in the sense that their 
parities are opposite. These two particles couple differently to particles 
with  other spins and parities, as will be shown in the next section, 
although they are produced by one and the same interaction.

Duality transformation turns the tensor 
propagator into a rescaled
pseudotensor one, with the opposite sign of the coupling constant:
\begin{eqnarray}
\beta {\tilde D}_{\rm T}^{\rm I}(G) &=&  
{- 2 \alpha\beta G 
\over{1 - 2 \left(\alpha - \beta \right)G\Pi_{\rm T}^{\rm I}(q)}} 
\nonumber \\
&-& 2 \beta \alpha G \left[
{1 \over{ 1 - 2 \left(\alpha - \beta \right) G \left(\Pi_{\rm T}^{\rm I}(q)
+ \Pi^{\rm II}(q)\right)}} - 
{1 \over{ 1 - 2 \left(\alpha - \beta \right) G \Pi_{\rm T}^{\rm I}(q)}} \right]
\nonumber \\
&=&  
{- 2 \beta \alpha G \over{ 
1 - 2 \left(\alpha - \beta \right) G \left(\Pi_{\rm T}^{\rm I}(q)
+ \Pi^{\rm II}(q)\right)}}
\nonumber \\
&=& 
{- 2 \beta \alpha G \over{1 + 
2 \left(\alpha - \beta \right) G \Pi_{\rm PT}^{\rm I}(q)}}
\nonumber \\
&=& - \alpha 
D_{\rm PT}^{\rm I}(G)~,
\label{e:dual4}
\end{eqnarray}
and similarly
\begin{eqnarray}
\beta {\tilde D}_{\rm T}^{\rm II}(G) &=&
- \alpha D_{\rm PT}^{\rm II}(G)~.
\label{e:dual5}
\end{eqnarray}
These imply the following identities
\begin{eqnarray}
\beta {\tilde D}_{\rm T}(G) &=& - \alpha D_{\rm PT}(G)
\nonumber \\
\alpha {\tilde D}_{\rm PT}(G) &=& - \beta D_{\rm T}(G) ~.
\label{e:dual6}
\end{eqnarray}
Inserting these results into Eq. (\ref{e:dual7}) we find
\begin{eqnarray}
D_{\rm T}^{\rm I~'} &=& 
{2 \left(\alpha - \beta \right) G_T \over{
1 - 2 \left(\alpha - \beta \right) G_T \Pi_{\rm T}^{\rm I}(q)}} 
\nonumber \\
D_{\rm T}^{\rm II~'} &=&  
2 \left(\alpha - \beta \right) G_T \left[
{1 \over{1 + 2 \left(\alpha - \beta \right) G_T \Pi_{\rm PT}^{\rm I}(q)}} - 
{1 \over{ 1 - 2 \left(\alpha - \beta \right) G_T \Pi_{\rm T}^{\rm I}(q)}} 
\right]
\nonumber \\
D_{\rm PT}^{\rm I~'} &=& 
{- 2 \left(\alpha - \beta \right) G_T \over{
1 + 2 \left(\alpha - \beta \right) G_T \Pi_{\rm PT}^{\rm I}(q)}} 
\nonumber \\
D_{\rm PT}^{\rm II~'} &=&  
- 2 \left(\alpha - \beta \right) G_T \left[
{1 \over{1 - 2 \left(\alpha - \beta \right) G_T \Pi_{\rm T}^{\rm I}(q)}} - 
{1 \over{ 1 + 2 \left(\alpha - \beta \right) G_T \Pi_{\rm PT}^{\rm I}(q)}} 
\right]~.
\label{e:prop3}
\end{eqnarray}
Hence we see that the total tensor and pseudotensor propagators are 
not identical, as some have conjectured, but are related to each other 
by the duality transformation.

As a check of our procedure we see that for a vanishing interaction 
Lagrangian (\ref{e:lag2}), i.e., with $\alpha = \beta$ the net propagation 
of the true (``total'') a.s. tensor, or a.s. pseudotensor modes also vanishes
\begin{eqnarray}
D_{\rm T}^{'}(\alpha = \beta) 
&=& 
D_{\rm T}^{'}(\alpha = \beta) = 0~,
\label{e:res1}
\end{eqnarray}
Moreover, as stated above, for tensor 't Hooft interaction 
Lagrangian (\ref{e:lag3}), i.e., with $\alpha \neq \beta$
the total T and PT mode propagators 
(\ref{e:res1}) are duality-gauge invariant as they depend only on the 
duality-gauge invariant linear combination $\alpha - \beta$.

\paragraph*{
Nambu-Goldstone bosons}
The poles in the propagators Eqs. (\ref{e:prop3}),
determine the masses of the T, PT states, while the 
residues determine their coupling constants to the quarks.
There are two sets of poles/masses
(a) pseudotensor
\begin{equation} 
D_{\rm PT}^{\rm I~'}(s) = 
{- 2 G_{\rm T} \over{1 + 
2 G_{\rm T} \Pi_{\rm PT}^{\rm I}(s)
}} 
= {g_{\rm PT}^{2} \over{
\left[(s + 6 m^{2}) F(s) + 2 m^2 \left(F(s) - 1\right) 
- m_{\rm T}^{2}\right]}}
\label{etap}
\end{equation}  
and (b) tensor
\begin{equation} 
D_{\rm T}^{\rm I~'}(s) = 
{2 G_{\rm T} \over{1 - 2 G_{\rm T} \Pi_{\rm T}^{\rm I}(s) }} 
= {
g_{\rm T}^{2} \over{
\left[(s - 6 m^{2}) F(s) + 2 m^2 \left(F(s) - 1\right) 
+ m_{\rm T}^{2}\right]}}~,
\label{vsigmap}
\end{equation} 
where we introduced the (gauge invariant) ``tensor mass'' $m_{\rm T}$ as
\begin{eqnarray}
m_{\rm T}^{2} = - {3 g_{p}^{2} \over 4 G_{\rm T}} ~,
\label{mth}
\end{eqnarray} 
and the associated zero-external-momentum coupling constants as
\begin{eqnarray}
g_{p}^{2} &=& 
\left({m \over f_{p}} \right)^{2} = 
{(2 \pi)^{2} \over{3}} 
\left(\sum_{s = 0}^{2} C_{s} \log(M^2_s/m^2) \right)^{-1}
\nonumber \\
&=& {2 \over 3} g_{\rm T}^{2} 
\left(1 + \left({g_{p}\over2\pi}\right)^{2}\right)
= {2 \over 3} g_{\rm PT}^{2} 
\left(1 + \left({g_{p}\over 2\pi}\right)^{2}\right),
\label{coupl}
\end{eqnarray}
where the $C_s$ and $M^2_s = m^2 + \alpha_s \Lambda^2$
are the standard parameters of the Pauli-Villars regularization
scheme \cite{iz80}, and $f_{p}$ is the p.s. ``pion'' decay constant.

We find a remarkable symmetry pattern in the mass spectrum: there 
are four poles
\footnote{Remember that there are two isospin channels, isoscalar 
and isovector, which differ only in the overall sign of the tensor 
coupling constant $G_T$.}
symmetrically placed about the origin
with locations at $\pm 6 m^{2} \pm m_{\rm T}^{2}$.
One finds two massless poles (in the chiral limit
\footnote{Some doubts have been expressed with regard to the NG nature of 
these massless poles. These doubts ought to be allayed by the fact that 
the T, PT states acquire a mass upon
explicit breaking of the chiral symmetry by current quark masses. 
This mass equals the pion mass under the same circumstances.}), 
one in the antisymmetric pseudotensor- and another 
in the a.s. tensor channel, at $s = 0$ (Nambu-Goldstone) 
provided that
\begin{equation}
m_{\rm T}^{2} 
= 6 m^{2} ~,
\label{e:mv} 
\end{equation}
holds,
which is equivalent to $G_{\rm T}^{-1} = - 8 f_{p}^{2} = 2 G_{\rm S}^{-1}$.
Note that precisely this ratio of the two coupling 
constants arises when one takes the Fierz transform of the
$N_f = 2$ 't Hooft interaction \cite{th76} as the interaction Lagrangian. 
The relations (\ref{mth}), (\ref{e:mv}) bear remarkable similarity to 
analogous relations for the vector mass and coupling constant in the 
ENJL model \cite{vd99}. In other words,
Eq. (\ref{e:mv}) defines a critical point in the space of a.s. T 
coupling constants 
in this theory. Change of $G_{\rm T}$ or $G_{\rm S}$ can lead to (phase) 
transitions to other phases of the theory, and thence to tachyons.

\section{The Higgs effect}
By replacing the partial derivative $\partial_{\mu}$ with the covariant one
$D_{\mu} = \partial_{\mu} - i e A_{\mu}~,$ 
in the Lagrangian (\ref{e:lag1}) and adding the gauge field Lagrangian
to it, we can couple a gauge field $A_{\mu}$ to the fermions in
this model. We will work in a class of covariant gauges parametrized by a 
gauge fixing parameter $\xi$. That amounts to adding the gauge-fixing term
\begin{eqnarray}
{\cal L}_{\rm gauge~fixing} = - {1 \over 2} \xi  
\left(\partial_{\mu} A^{\mu}\right)^{2}  
\label{e:gf}
\end{eqnarray}
to the Lagrangian Eq. (\ref{e:lag1}) and consequently having 
\begin{eqnarray}
{D}^{\mu \nu}(q) = {- 1 \over q^{2}} \left[g^{\mu \nu} - \left(1 - 
{1 \over \xi}\right)  {q^{\mu} q^{\nu} \over q^{2}} \right]  
\label{e:photon}
\end{eqnarray}
as the ``bare" gauge boson propagator. 
This propagator is ``dressed" by vacuum polarization correction parametrized
by the gauge invariant tensor
\begin{eqnarray}
{\pi}_{\mu \nu}(q) = 
\left(q_{\mu} q_{\nu} - g_{\mu \nu} q^{2} \right) \pi (q^{2})
\label{e:pol}
\end{eqnarray}
according to the Schwinger-Dyson equation
\begin{eqnarray}
{\bf D}^{\mu \nu}(q) = {D}^{\mu \nu}(q) +  {D}^{\mu \lambda}(q) 
{\pi}_{\lambda \sigma}(q)
{\bf D}^{\sigma \nu}(q) ~. \
\label{e:SDE}
\end{eqnarray}
The solution to this SDE reads 
\begin{eqnarray}
{\bf D}^{\mu \nu}(q) = {- 1 \over q^{2}} \left[
\left(g^{\mu \nu} - {q^{\mu} q^{\nu} \over q^{2}} \right) 
{1 \over{1 - \pi (q)}}  
+ {1 \over \xi} {q^{\mu} q^{\nu} \over q^{2}} \right]  ~.
\label{e:dressedphoton}
\end{eqnarray}
Schwinger observed \cite{sch61} that when the vacuum polarization function
$\pi(q^2)$ has a simple pole at $q^2 = 0$, the dressed gauge boson propagator 
\begin{eqnarray}
q^{2} \left(1 - \pi (q)\right) = q^{2} - M_{\rm V}^{2}~,
\label{e:pole}
\end{eqnarray}
acquires a {\it finite gauge-invariant dressed mass} $M_V$
determined by the residue of $\pi$ at the pole as follows
\begin{eqnarray}
M_{\rm V}^{2} = \lim_{q^{2} \to 0} q^{2} \pi (q)~. 
\label{e:mass}
\end{eqnarray}
This is also the way the ``conventional'' Higgs mechanism operates
\cite{higgs64}.
We see in Eq. (\ref{e:pole}) that it is absolutely crucial for $\pi (q)$ to 
have a pole at precisely $q^{2} = 0$, or else the dressed gauge boson remains 
massless.

\begin{figure}
\begin{center}
\epsfig{file=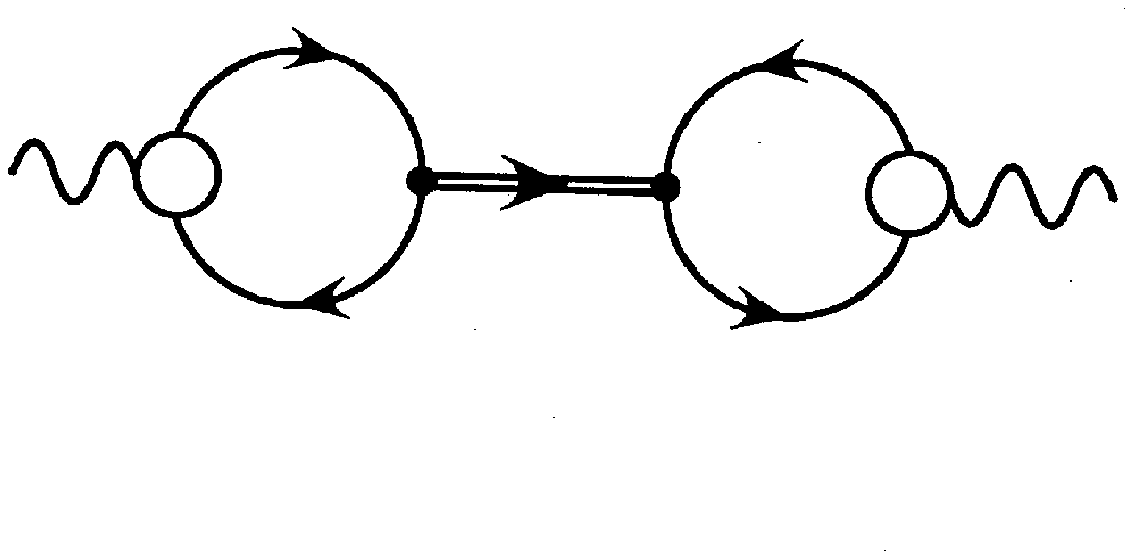,width=10cm} 
\end{center} 
\caption{
The vector current polarization tensor $\pi_{\mu \nu}$. 
The wavy lines are the Yang-Mills vector bosons,
solid lines are constituent quarks and the double solid line is the 
antisymmetric pseudo-tensor boson.}
\label{f:2}
\end{figure}

That this is indeed the case in
the present theory, we can see by constructing the 
vector polarization tensor $\pi_{\mu \nu}$, see Fig. \ref{f:2}. 
For that we need the vector-pseudotensor [V-PT] transition matrix element
$\Pi_{\mu \nu \alpha}^{V-PT}$, which is again given by the simple one-loop 
graph appearing in Fig. \ref{f:2}. 
One finds
\begin{eqnarray}
\Pi_{\mu \nu \alpha}^{\rm V-PT}(s) &=& 
i e m g_{p}^{-2} F(s) \varepsilon_{\mu \nu \alpha \beta} q^{\beta}~.
\label{e:pi2}
\end{eqnarray}
[Note that the analogous A-T transition tesnor vanishes, i.e., 
a.s. tensors do {\it not} couple to axial-vector currents. This shows 
a definite asymmetry between the two sectors with opposite parities.]
Inserting this into 
\begin{eqnarray}
\pi_{\mu \nu}(q) &=& 
\Pi_{\mu \alpha \beta}^{V-PT}(q) D_{PT}(q^2) \Pi_{\nu \alpha \beta}^{PT-V}(q)
\nonumber \\
&=&
\left[q_{\mu} q_{\nu} - q^2 g_{\mu \nu}\right]
\left(e m g_{p}^{-2}\right)^{2} 
{3 g_{p}^{2} F^{2}(q^2) \over{
\left[(q^2 + 6 m^{2}) F(q^2) + 2 m^2 \left(F(q^2) - 1\right) 
- m_{\rm T}^{2}\right]}}~,
\label{e:pi3}
\end{eqnarray}
we find 
\begin{eqnarray}
M_{\rm V}^{2} = \lim_{q^{2} \to 0} q^{2} \pi (q) = 
{3 \left(e f_{p}\right)^{2} \over{1 + \left({g_{p}\over{2 \pi}}\right)^{2}}}
\label{e:mass1}
\end{eqnarray}
in the $m_{\rm T} \to \sqrt{6} m$ limit.
In other words, one must have $4 G_{S} = 8 G_T = - f_{p}^{-2}$ for the 
Higgs mechanism to be operative, the same 
condition as for the masslessness of 
the antisymmetric pseudotensor NG bosons. 

\section{Discussion and conclusions}

In conclusion, we have shown that: 
1) a non-Abelian symmetry model with dynamical symmetry breaking of the NJL 
type and an antisymmetric tensor fermion self-interaction 
leads to massless composite antisymmetric tensor NG bosons at tensor coupling 
$G_T = {1\over 2} G_S$; and 
2) vector gauge bosons coupled to this system acquire a mass of 
${\sqrt{3} e f_{p} \over \sqrt{1 + \left({g_{p}\over{2 \pi}}\right)^{2}}}$,
where $e$ is the gauge coupling constant and $f_{p}$ is the scalar 
(Higgs) v.e.v.. Such a result may be termed an ``antisymmetric tensor 
Schwinger-Higgs mechanism'' with composite a.s. T states. 
The distinction from the usual (scalar)
Higgs mechanism is that there is no a.s. T Higgs particle. 
A similar effect with elementary a.s. T fields has been recognized
by St\" uckelberg \cite{st38}. 

Analogous Higgs mechanism for {\it axial-vector} gauge fields with the 
original NJL model has been discussed by Freundlich and Luri\' e \cite{fl70}.
The Freundlich-Luri\' e scheme
is the dynamical symmetry breaking mechanism currently used in many
``top-condensation'' models of the electroweak interactions \cite{cvet97}.
The parity of the would-be NG boson (``the Higgs-Kibble ghost'') is 
unimportant in applications to electroweak interactions, 
where the gauge fields are one part vector and one part axial-vector, 
but it is crucial in applications to QCD, which theory conserves parity
and whose quanta (gluons) are vector particles. 
Phenomenological consequences of an a.s. tensor Higgs mechanism 
as applied to the Salam-Weinberg model remain to be worked out.
Applications to the confinement problem in QCD are being worked on. 
Last, but not least, this new a.s tensor Higgs mechanism may have 
applications in hadronic effective theories \cite{weise90}.




\paragraph*{Acknowledgements}
I would like to thank A. Tetervak for giving me access to his Dirac
trace evaluation program TamarA.

\end{document}